\documentclass{aa}
\usepackage{lscape,graphicx}
\usepackage{./natbib}
\usepackage{./rotating}
\bibpunct{(}{)}{;}{a}{}{ }

\begin{document}

\title{ Tracing a Z-track in the M31  X-ray binary RX\thinspace J0042.6+4115}
\author{ R. Barnard\inst{1}
    \and U. Kolb\inst{1}
    \and J. P. Osborne\inst{2}}

\offprints{R. Barnard, \email{r.barnard@open.ac.uk}}

\institute{ The Department of Physics and Astronomy, The Open University, Walton Hall, Milton Keynes, MK7 6BT, U.K.
     \and The Department of Physics and Astronomy, The University of Leicester, Leicester, LE1 7RH, U.K.}
\date{ Received 27 May 2003 / Accepted 16 September 2003}

\abstract{
Four XMM-Newton observations of the core of M31, spaced at 6 month intervals, show that the brightest point X-ray source, RX\thinspace J0042.6+4115, has a 0.4--10 keV luminosity of $\sim$5  10$^{38}$ erg s$^{-1}$, and exhibits significant  variability in intensity and X-ray spectrum over a time scale of $\sim$100 s including hard flares; such behaviour is only observed in Z-sources and transient black hole binaries in our Galaxy. The lightcurves, X-ray spectra and hardness-intensity data from the four XMM-Newton observations all strongly suggest that it is a Z-source, bringing the total number of known Z-sources to nine.
 } 

\titlerunning{A Z-source candidate in M31}
\maketitle

\section{Introduction}
\label{intro}
The Andromeda galaxy (M31) is the nearest spiral galaxy, at a distance of 760 kpc \citep{vdb00}. The X-ray emission from M31 is dominated by point sources.
While the study of variability in X-ray sources in external galaxies has been limited by the sensitivity of observatories prior to XMM-Newton, its three large X-ray telescopes can make variability within a single observation  visible down to fluxes of $\sim$10$^{-14}$ erg cm$^{-2}$  s$^{-1}$ in the 0.3 -- 10 keV band \citep[e.g.][]{jan01}, equivalent to luminosities of $\sim$10$^{36}$ erg s$^{-1}$ in M31. The XMM-Newton observations of M31 have already revealed a persistent, stellar-mass black hole binary \citep[][ Paper 1]{bok03}, periodic intensity dips in the M31 globular cluster X-ray source Bo158 \citep{tru02} and pulsations in a white-dwarf transient \citep{osb01}. 

RX J0042.6+4115, the subject of this paper,  was discovered in the Einstein survey of \citet{tf91}, and was named by \citet{S97}, who associated it with  a globular cluster. However, its location was more accurately given in a recent Chandra survey \citep{K02} as 00$^{\rm h}$~42$^{\rm m}$~38$\fs$581~+41$\degr$~16$\arcmin$~03$\farcs$80 with a positional error of 0$\farcs$01, and was not associated with any foreground object, globular cluster or background AGN. \citet{K02} found variability in the lightcurves and spectra of RX J0042.6+4115 and report a luminosity in the 0.3--7 keV band of 1.5 10$^{38}$ erg s$^{-1}$, assuming a standard absorbed power law spectral model with an index of 1.7 and absorption equivalent to 10$^{21}$ hydrogen atom cm$^{-2}$. Our results from four XMM-Newton observations of RX J0042.6+4115 reveal  0.3--7 keV luminosities of  $\sim$3--4 times that reported by \citet{K02} while   spectral fits  required a blackbody component in addition to a power law component. Furthermore, significant variability is observed over time-scales of $\sim$100 s. Of the Galactic X-ray binary population, such behaviour is only observed in Z-sources and X-ray transients. As this object is clearly not a transient, we argue that RX\thinspace J0042.6+4115 is the 9th Z-source.

The paper is constructed as follows: the properties of Z-sources are briefly reviewed in the next  section;  the XMM-Newton observations and analysis techniques are then discussed;  lightcurves, colour-colour diagrams  and energy spectra are presented in the Results section and the case for classifying  RX\thinspace J0042.6+4116 as an analogue to the Galactic Z-sources presented in the Discussion section.

\section{Properties of Z-sources}
\begin{figure}[!t]
\resizebox{\hsize}{!}{\includegraphics{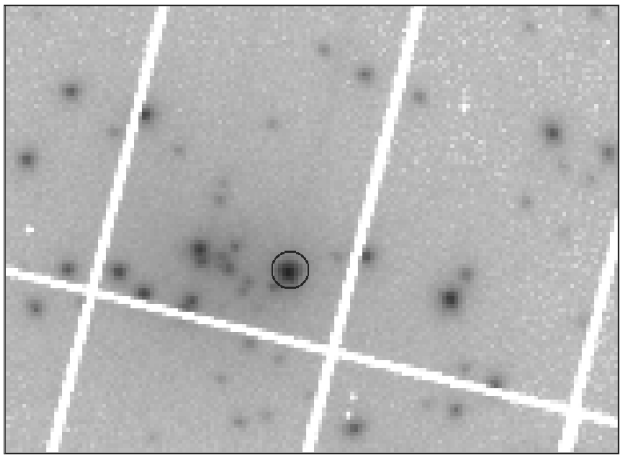}}
\caption{ A detail of the 0.3--10 keV PN image from the June 2001  XMM-Newton observation of the core of M31; north is up, east is left. The image is log scaled and 5$\arcmin$ across.  RX\thinspace J0042.6+4115 is circled in black, the circle defining the source extraction region.}\label{obx3}
\end{figure}
The Z-source subclass of low mass X-ray binaries (LMXB) is characterised by high X-ray luminosities  \citep[$\geq$10$^{38}$ erg s$^{-1}$,][]{hv89} and  intensity variations of up to 100\% over timescales of minutes \citep[][ and references within]{wms76}; they are named after the patterns traced on their colour-colour diagrams. If X-ray lightcurves of a source are obtained in several energy bands, then one can define hard and soft colours to see spectral variations over time; the commonly used  colour-colour diagram (CD) is a plot of hard colour vs. soft colour. Z-sources exhibit three-branched, approximately Z-shaped CDs, with temporal properties that are strongly correlated to the position of the source on the CD \citep{hv89}. The branches are historically named the horizontal branch (HB), normal branch (NB) and flaring branch (FB). During spectral evolution, the Z-source moves smoothly along the branches without jumping from one branch to another; the  mass accretion rate is thought to increase as the source travels in the sense HB to NB to FB \citep{hv89}. The Z-sources exhibit significant branch movement over time-scales of $\sim$100 s or less and trace out their complete pattern in $\sim$1 day \citep{mrc02}.
All Galactic Z-sources exhibit kilohertz quasi-periodic oscillations \citep[see][ and references within]{lvv01}. This tells us that they all contain neutron star primaries, since the power density spectra of LMXB with black hole primaries are not thought to contain any power at $>$500 Hz \citep{sr00}.
These Z-sources may be divided further into two subsets: \object{Cygnus\thinspace X-2}, \object{GX\thinspace 5$-$1} and \object{GX\thinspace 340+0} have strong, horizontal HBs and weak flaring branches while \object{Sco\thinspace X-1}, \object{GX\thinspace 17+2} and \object{GX\thinspace 349+2} have strong FBs and weak, vertical HBs \citep{hv89}. The underlying differences that account for this division of Z-sources is unknown. Two additional Z-source candidates have recently been identified: \object{Circinus X-1} \citep[e.g.][]{iar01} and the first extra-galactic Z-source \object{LMC\thinspace X-2} \citep{sk00}.  
It is highly likely that the set of 6--7 known Z-sources in our Galaxy is complete, due to their high X-ray luminosities, yet they account for only $\sim 5\%$ of the Galactic LMXB \citep{lvv01}.

The spectral evolution of Sco\thinspace X-1 along its flaring branch was studied by \citet{bcb03}, using data from the Rossi-XTE observatory \citep{brs93}; luminosities  of $\sim$5--12 10$^{38}$ erg s$^{-1}$ were observed in the 1--30 keV energy band. This exceeds the Eddington luminosity for an accreting,  1.4 M$_{\odot}$ neutron star, the luminosity where the radiation pressure on accreting material balances the gravitation forces ($\sim$2 10$^{38}$ erg s$^{-1}$), \citep{fkr92}. However, the majority of the X-ray emission comes from an extended  accretion disc corona and only 10--30\% of the 1--30 keV  emission comes from the neutron star surface, in the form of blackbody emission; the luminosity of the neutron star is generally sub-Eddington \citep{bcb03}. We can therefore expect the X-ray spectrum of an analogue of Sco\thinspace X-1 to contain a strong blackbody component as well as the standard power law component fitted by \citet{K02}.

\section{Observations}

Four XMM-Newton observations were made of the core of M31; data from the three EPIC detectors in each observation  were analysed for variation between observations and variability within observations. The three imaging CCD instruments, two MOS \citep{turn01} and one PN \citep{stru01} share a common, circular field of view with a radius of 30$\arcmin$ \citep{jan01}. A journal of the XMM-Newton observations of the core of M31 are presented in Table~\ref{journ}.
\begin{table}[!t]
\centering
\caption{Journal of XMM-Newton observations of the \object{M31} core}\label{journ}
\begin{tabular}{lllll}
\noalign{\smallskip}
\hline
\noalign{\smallskip}

Observation & Date & Exp  & Filter\\
\noalign{\smallskip}
\hline
\noalign{\smallskip}
1 &  25/07/00 (rev0100)& 34 ks& Medium \\
2 & 27/12/00 (rev0193)& 13 ks&  Medium\\
3 & 29/06/01 (rev0285)& 56 ks &Medium \\
4 & 06/01/02 (rev0381)& 61 ks&  Thin\\
\noalign{\smallskip}
\hline
\noalign{\smallskip}

\end{tabular}
\end{table}
Lightcurves of RX\thinspace J0042.6+4115 were obtained in the 0.3--2.5, 2.5--10, 4.0--10 and 0.3--10 keV energy bands from the PN and MOS detectors at their highest time resolutions. The analysis closely follows that performed in Paper  1, with two minor differences. Firstly, the background contribution to the lightcurves was very small, and all good data were used. 
\begin{table*}[!t]
\caption{Variability of EPIC lightcurves of RX\thinspace J0042.6+4115 in different energy bands with 400 s binning: $\chi^2$/dof and P (the probability that a random fit to the data would be a worse fit) are  shown for the best fit lines of constant intensity to lightcurves in three energy bands for each observation.}\label{var}
\begin{tabular}{lllllll}
\noalign{\smallskip}
\hline
\noalign{\smallskip}
Observation & $\chi^{2}_{\rm 0.3-2.5 keV}$/dof &  P$_{\rm 0.3-2.5 keV}$ & $\chi^{2}_{\rm 2.5-10 keV}$/dof&   P$_{\rm 2.5-10 keV}$& $\chi^{2}_{\rm 0.3-10 keV}$/dof &  P$_{\rm 0.3-10 keV}$\\
\hline
\noalign{\smallskip}
1 & 100/77 &  0.04  & 143/77 & 7 10$^{-6}$  & 162/77 & 6 10$^{-6}$\\
2 & 33/24 & 0.10 & 23/24 & 0.52  & 28/24 & 0.26\\
3 & 513/131 & $\ll$10$^{-6}$ &  412/131 &   $\ll 10^{-6}$ & 259/131 &$\ll$10$^{-6}$ \\
4 & 730/152 &  $\ll$10$^{-6}$ &  497/152 & $\ll$10$^{-6}$ & 441/152 &  $\ll$10$^{-6}$ \\
\noalign{\smallskip}
\hline
\noalign{\smallskip}
\end{tabular}
\end{table*}
\begin{figure}[!t]
\resizebox{\hsize}{!}{\includegraphics{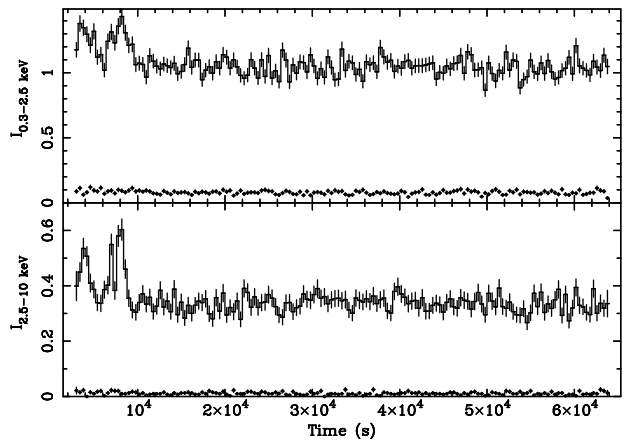}}
\caption{ The EPIC (PN + MOS1 +MOS2)  lightcurves of RX\thinspace J0042.6+4115 from observation 4 in the 0.3--2.5 keV (top panel) and 2.5--10 keV (bottom panel) energy bands; the upper curve in each panel is the background-subtracted source curve, while  the lower curve is the background curve.  The lightcurves were accumulated with time bins of 400 s. Clearly, the amplitude of flaring is higher in the 2.5--10 keV energy band, and it is also clear that the background makes no contribution to the flares. }\label{hlcomp}
\end{figure}
 Secondly, spectra from the PN only were analyzed; the data were accumulated into 250 eV bins and bins below 0.4 and above 10 keV were ignored. 

\section{Results}

 Observations 1, 3 and 4 exhibit significant variability in all energy bands, while the lightcurves from observation 2 exhibit no significant variability; each lightcurve was fitted with a line of constant intensity and the resulting best fit $\chi^2$/dof and probability of randomly getting a larger $\chi^2$/dof (P) are presented in Table~\ref{var}. Strong intensity enhancements occur at the start of observations 3 and 4; in fact, the lightcurves of RX\thinspace J0042.6+4115 closely resemble the 1998, January RXTE lightcurves of Sco\thinspace X-1, which captured a transition between the FB and the NB \citep{bcb03}. The flares in Sco\thinspace X-1 last for a few thousand seconds and are hard, meaning that the fractional  amplitude of the flares increases with energy  \citep{wms76}. Combined EPIC (PN + MOS1 + MOS2) lightcurves of RX\thinspace J0042.6+4115 from observation 4 in the 0.3--2.5 and 2.5--10 keV energy bands are given in Fig.~\ref{hlcomp}   with 400 s binning; clearly the relative amplitude of flaring is higher in the 2.5--10 keV energy band ($\sim$100\%) than in the 0.3-2.5 keV band ($\sim33$\%).

\begin{figure}[!t]
\resizebox{\hsize}{!}{\includegraphics{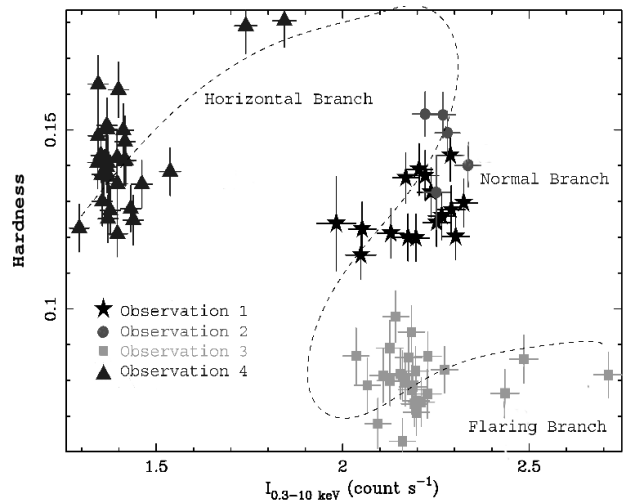}}
\caption{ Hardness ratio vs. intensity for 4 XMM-Newton observations of RX\thinspace J0042.6+4115; the hardness ratio is defined as I$_{\rm 4.0-10 keV}$/I$_{\rm 0.3-2.5 keV}$ and the data is binned to 2000 s. The dashed line shows a possible Z-track for RX\thinspace J0042.6+4115. }\label{hid}
\end{figure}

 Hardness-intensity data was obtained for the 4 observations of RX\thinspace J0042.6+4115; the hardness was defined as the ratio of the 4.0--10 keV intensity to the 0.3--2.5 keV intensity, and data were integrated over 2000 s. 
The high energy band 4.0--10 keV was chosen to better separate the three branches; Z-sources are more variable at higher energies than at low energies \citep[see e.g.][ who modelled the X-ray spectrum as bremsstrahlung emission and interpreted the increase in hardness with intensity as an increase in temperature]{wms76}. 
Fig.~\ref{hid} shows the hardness vs intensity diagram (HID) for all observations of RX\thinspace J0042.6+4115; stars, circles, triangles and squares represent data from observations 1, 2, 3 and 4 respectively. The resulting HID appears to show fragments of the Z-pattern characteristic of the Cygnus\thinspace X-2 like Z-sources \citep{hv89}; it appears that
observation 1 is on the NB, observation 2 is near the HB/NB apex,
observation 3 is on the FB, and observation 4 is on the HB. 

\begin{figure}[!t]
\resizebox{\hsize}{!}{\includegraphics{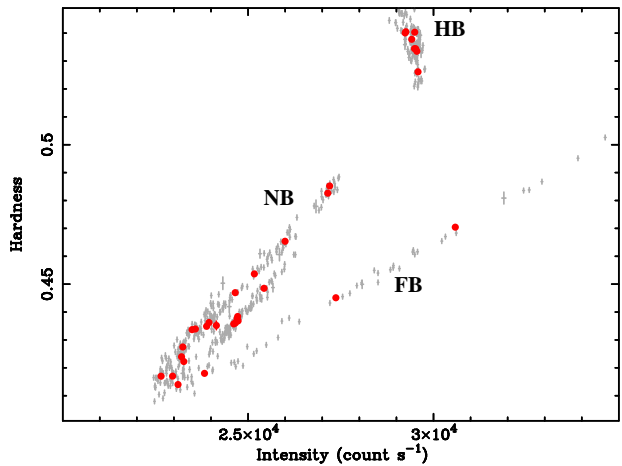}}
\caption{ Hardness vs. intensity data from RXTE observations of the Galactic Z-source Sco\thinspace X-1 with 96 s binning (grey crosses) and 2000 s binning (circles). The HID at 2000 s binning clearly reflects the branch movement, even though the HID at 96s demonstrates that spectral variability occurs over much shorter time-scales.}\label{hicomp}
\end{figure}

In order to illustrate the effects of long integration times on HID for Z-sources, we present in Fig.~\ref{hicomp} hardness vs. intensity data from  400 ks of RXTE data on Sco\thinspace X-1 using 96 s binning and 2000 s binning. We see that using 2000 s bins preserves the shape of the HID that would be seen at  96 s resolution, although the curve is more sparsely sampled; hence we can be confident that our HID of RX\thinspace J0042.6+4115 is consistent with Z-source HID. 

\begin{figure}
\resizebox{\hsize}{!}{\includegraphics[angle=270]{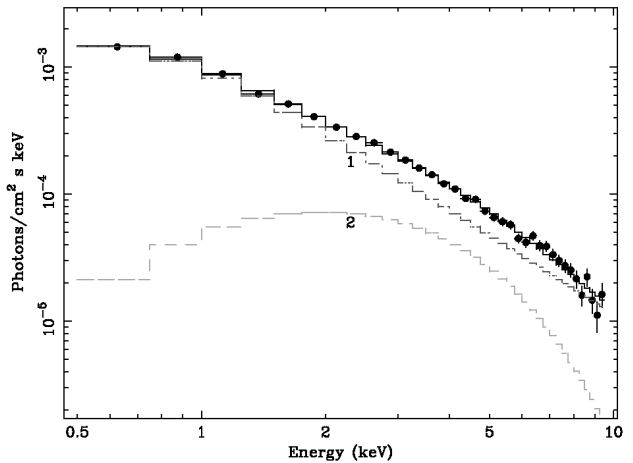}}
\caption{The PN spectrum of RX\thinspace J0042.3+4115 from observation 1; the data (filled circles) are fitted with the sum of two components, a power law component (1) and a blackbody component (2). }\label{spec}
\end{figure}

The intervals of high intensity variability were omitted from spectral analysis, because the hardness ratios of these data (and therefore spectra) significantly varied; as a result, the spectrum from observation 1 had an exposure of only 10 ks, while the spectra of observations 3 and 4 both had exposures of $\sim$50 ks. The PN  spectra from the four observations of RX\thinspace J0042.6+4115 were first fitted by simple power law models (A1--A4). None of these fits were statistically acceptable.  Fits to the spectrum of observation 4 using models consisting of a single bremsstrahlung component ($\chi^2$/dof=103/29) or a single  {\sc comptt} component ($\chi^2$/dof = 247/27) were also rejected.  Hence a two component emission model was applied, consisting of a blackbody and a power law. These two component models (B1--B4) all provided good fits to their respective spectra.  All  models incorporated freely-fitted, line-of-sight absorption. The  unfolded, 0.4--10 keV PN spectrum of RX\thinspace J0042.6+4115 from observation 1 is presented in Fig.~\ref{spec}; the data are shown (filled circles) along with the power law (1) and blackbody (2) components, and the sum for the best fit model.  The best fit parameters, $\chi^2$/dof and fluxes of each model are presented in Table~\ref{specfit}; 90\% confidence errors were obtained for the two-component models, but not the single component models since the fits were unacceptable.
The spectra were then tested for the presence of line emission. There is evidence for a 6.4 keV line, corresponding to fluorescent iron K$\alpha$ emission in observation 3 only: F-testing showed that the improvement in $\chi^2$/dof due to fitting the line had a 3\% probability of being random.

\begin{table*}[!t]
\caption{Results from fitting the 0.3--10 keV PN X-ray spectra of RX\thinspace J0042.6+4115 with spectral models A[n] and B[n]; A1--A4 are single component power law models fitted to observations 1--4 respectively, and B1--B4 are equivalent two-component models, consisting of a blackbody and a power law. Best fit parameters are given, along with the $\chi^2$/dof and unabsorbed 0.4--10 keV fluxes of the total emission and the blackbody component, where applicable. Numbers in parentheses indicate 90\% confidence errors on the last digit. }\label{specfit}
\centering
\begin{tabular}{llllllllllll}
\noalign{\smallskip}
\hline
\noalign{\smallskip}
Model & $N_{\rm H}$ /10$^{22}$& k$T$ & $n_{\rm BB}$ /10$^{34}$ & $\Gamma$ & $n_{\rm PO}$/10$^{-3}$ & $\chi^2$/dof & $F^{\rm tot}/10^{-12}$& $F^{\rm BB}/10^{-12}$\\
 & atom cm$^{-2}$ & keV & erg s$^{-1}$ & & photon cm$^{-2}$ & & erg cm$^{-2}$ s$^{-1}$  & erg cm$^{-2}$ s$^{-1}$ \\
 & & &at 10 kpc & & at 1 keV\\

\noalign{\smallskip}
\hline
\noalign{\smallskip}
A1 & 0.14 & $\dots$ & $\dots$ &1.9 & 1.4  & 49/29 & 8.1  & $\dots$\\
A2 & 0.15 & $\dots$ & $\dots$ & 1.8& 1.4  & 49/24 & 8.0  & $\dots$ \\
A3 & 0.14 & $\dots$ & $\dots$ & 1.8 & 1.2  & 351/30& 7.3  &$\dots$\\
A4 & 0.14 & $\dots$ & $\dots$ & 1.8 & 1.16 & 258/29 & 7.2 & $\dots$\\
B1 & 0.14(3) & 1.08(10) & 1.28(18)  & 2.02(3) & 1.36(3)  & 34/27 & 8(3) & 1.1(4) \\ 
B2 & 0.14(2) & 1.05(8) & 1.87(15)  & 1.95(4) & 1.22(3) & 23/22 & 8(2) & 1.5(5) \\
B3 & 0.16(2) & 1.11(3) & 2.79(8)  & 2.24(2) & 1.12(1)  & 37/28 & 7.3(8)  & 2.3(3) \\
B4 & 0.13(2) & 1.08(3) & 2.65(10)  & 2.10(3) & 1.00(2)  & 33/27 & 7.0(9)  & 2.2(3) \\
\noalign{\smallskip}
\hline
\noalign{\smallskip}
\end{tabular}
\end{table*}

The spectral shape parameters from the two component models are fairly consistent from observation to observation, with line-of-sight absorption of $\sim$0.14 10$^{22}$ H atom cm$^{-2}$, a blackbody temperature equivalent to kT $\sim$1.1 keV, and a photon index of $\sim$2.1. However,  the normalisations of the two emission components change significantly. This is entirely consistent with the spectral evolution of Sco\thinspace X-1 \citep{bcb03}. The total 0.4--10 keV luminosity ranges over 4.8--5.6 10$^{38}$ erg s$^{-1}$, or $\sim$2.5 times the Eddington limit for a 1.4 M$_{\odot}$ neutron star, while the blackbody contribution is  0.7--1.6 10$^{38}$ erg s$^{-1}$, which is sub-Eddington. The total 0.4--10 keV luminosity decreases by $\sim$10\% over the 4 observations, while the blackbody luminosity increases by 100\%. The peculiar softness of observation 3 data appears to be due to the steep photon index, bearing out the association of observation 3 data with the flaring branch -- the photon index steepens along the flaring branch in Sco\thinspace X-1 \citep{bcb03}. Meanwhile the hardness of observation 4 data may be explained by the  increased blackbody contribution to the X-ray flux, since its fractional contribution to the 4.0--10 keV flux is greater than for the 0.3--2.5 keV flux, as demonstrated in Fig.~\ref{spec}.

\section{Discussion}

In order to place RX\thinspace J0042.6+4115 in M31, we must show that it is not a foreground object or a background AGN. 
 If RX\thinspace J0042.6+4115 was local, it would have to be within a few kpc, and so the 0.4--10 keV luminosity would be reduced by a factor of $\sim$10$^6$, to $\sim$5 10$^{32}$ erg s$^{-1}$. Thus RX\thinspace J0042.6+4115 would be too faint to be a persistently bright LMXB \citep{lvv95}  and would be either a cataclysmic variable or soft X-ray transient (SXT)  in quiescence.
However, the observed X-ray spectra for RX\thinspace J0042.6+4115 are completely unlike any observed in cataclysmic variables  \citep[see ][ and references within]{kn03}. Also the blackbody temperatures in our spectra are many times higher than expected for a  quiescent SXT \citep[c.f. $\sim$ 0.3 keV, ][]{mm01}.  Finally, they are inconsistent with observed spectra from AGN, which in the 0.5--10 keV energy range are described by a simple power law \citep{page98} whereas the spectra of RX\thinspace J0042.6+4115 require  a  second component.  It would be hard to understand this object if it were in our Galaxy or in a distant
background galaxy, whereas it is easily comprehensible as a member of M31.

RX\thinspace J0042.6+4115 is therefore a high luminosity X-ray binary, with a total 0.4--10 keV luminosity of 5--6 10$^{38}$ erg s$^{-1}$, including  0.8--1.7 10$^{38}$ erg s$^{-1}$ from a $\sim$1 keV blackbody component. The total luminosity exceeds the Eddington limit for a 1.4 M$_{\odot}$ neutron star, however the blackbody component does not. Two Galactic Z-sources are known to regularly exceed the Eddington limit: Sco\thinspace X-1 and GX\thinspace 5$-$1 \citep[e.g.][]{lvv01}, while of the black hole XB, only transients in outburst \citep{csv97} or  extreme oddballs such as  GRS\thinspace 1915+105 \citep[e.g.][]{bel00} reach the   luminosities we have observed in RX\thinspace J0042.6+4115. 

Outbursts in X-ray transients generally rise in intensity quickly (a few days) and decay exponentially; the decline time-scales cluster around 30 days, although time-scales of 400 days are known \citep{csv97}. The first detection of RX\thinspace J0042.6+4115 was over 20 years ago by the Einstein satellite, when the 0.2-4.0 keV luminosity was 2.4 10$^{38}$ erg s$^{-1}$ \citep{tf91}, and it has been persistently bright in every observation since \citep[e.g.][]{S97,S01}, and so is highly  unlikely to be a transient.

Since the three branches of the Z were observed over 18 months, it is possible that RX\thinspace J0042.6+4115 is a very luminous atoll source. However, \citet{mrc02} report that three-branched colour diagrams in atoll sources  are only observed in systems where the intensity varies by a factor of $\geq$80;  atoll sources that vary in X-ray intensity by no more than a factor of 10 only  ever display a portion of the pattern \citep{mrc02}. The 0.3--10 keV intensity of RX\thinspace J0042.6+4115 varies by a factor of $\sim$2 over the four XMM-Newton observations, and yet three patterns of hardness-intensity behaviour are observed. In this regard, RX\thinspace J0042.6+4115 is unlike any Galactic atoll source.

 The lightcurves, hardness-intensity diagram, and the luminosities and composition of the X-ray spectra all strongly suggest that RX\thinspace J0042.6+4115 is a Z-source; its luminosity and Z-pattern suggest that it is a super-Eddington, Cyg\thinspace X-2-like Z-source analogous to GX\thinspace 5$-$1. 
If RX\thinspace J0042.6+4115 is a Z-source, it would be only the 9th to be found.

\begin{acknowledgements}
 We thank the referee for providing useful suggestions for improving the paper. This work is supported by PPARC. Data from the HEASARC public data archive for RXTE were used in constructing Fig.~\ref{hicomp}. 
\end{acknowledgements}

\bibliographystyle{./bibtex/aa}
\bibliography{m31}
\end{document}